\documentclass{iaus278}
\usepackage{graphicx}
\pubyear{2011}
\volume{278}  
\pagerange{65--73}
\jname{``Oxford IX" International Symposium on Archaeoastronomy}
\editors{Clive L.N. Ruggles, ed.}

\title[The Tom\'ar\^aho conception of the sky]                   
{The Tom\'ar\^aho conception of the sky} 

\author[Guillermo Sequera \& Alejandro Gangui]                 
{Guillermo Sequera$^{1}$ \and Alejandro Gangui$^{2,3}$}        

\affiliation{
$^1$Secretar{\'\i}a Nacional de Cultura de Paraguay \\[\affilskip]  
$^2$Instituto de Astronom{\'\i}a y F{\'\i}sica del Espacio (CONICET--UBA),
\\ Ciudad Universitaria, 1428 Buenos Aires, Argentina
\\[\affilskip]
$^3$Centro de Formaci\'on e Investigaci\'on en la Ense\~nanza de las Ciencias, \\ FCEyN, Universidad de Buenos Aires, Argentina
\\ email: {\tt gangui@df.uba.ar} 
}

\begin{document}
\maketitle

\begin{abstract}

The small community of the Tom\'ar\^aho, an ethnic group culturally derived from the Zamucos, have become known in the South American and world anthropological scenario in recent
times. This group, far from the banks of the Paraguay river, remained concealed from organized modern societies for many years. Like any other groups of people in close
contact with nature, the Tom\'ar\^aho developed a profound and rich world-view which parallels other more widely researched aboriginal cultures as well as showing distinctive
features of their own. This is also apparent in their imagery of the sky and of the characters that are closely connected with the celestial sphere. 

This paper is based on the lengthy anthropological studies of G. Sequera. We have recently undertaken a project to
carry out a detailed analysis of the different astronomical elements present in the imagined sky of the Tom\'ar\^aho and other Chamacoco ethnic groups.  We will briefly review
some aspects of this aboriginal culture: places where they live, regions of influence in the past, their linguistic family, their living habits and how the advancement of
civilization affected their culture and survival. We will later mention the fieldwork carried out for decades and some of the existing studies and publications. We will also
make a brief description of the methodology of this work and special anthropological practices. Last but not least, we will focus on the Tom\'ar\^aho conception of the sky and describe the research work we have been doing in recent times.

\end{abstract}

\firstsection 

\section{Tom\'ar\^aho culture}

The act of living in a certain culture may be displayed in many different ways. One of these is creativity, at stake when observing the changing phenomena of the
surrounding natural world: giving names to things; naming people, animals and plants; in short, building up the knowledge and symbols to enable them to acquire an identity and survive
as a society. The present-day groups aspire to this, completely paralleling the cultures in the past, for whom this was a strong motivation---both of the more technologically developed groups as
well as the ones who did not follow the frantic development of `civilization' and remained in closer contact with nature. The small community of the Tom\'ar\^aho is an example of
the latter.

A first contact with members of this ethnic group took place in 1986 when one of the authors (Sequera) was able to see for himself the situation of near slavery and strong
dependence the members of this people had to put up with in their relationship with the Carlos Casado company, which was industrializing the region. This Anglo-Argentinian enterprise, in complicity with the Paraguayan government since the end
of the 19th century (a few years after the end of the War of the Triple Alliance which weakened the nation considerably), had
illegally taken the lands of the Tom\'ar\^aho and exploited the quebracho woods extensively in order to obtain timber for railway sleepers and to stock their tanning
plants. With no recognized rights of any kind, the natives were made to work as axemen under hard labour conditions and to be, in their turn, witnesses to the plunder of the
Chaco woods. In this way they became helpless victims of the subjugation of the environment that was vital to them (Sequera 2006). 

Given the importance of carrying out an anthropological study of this ethnic minority, comparative studies were undertaken of ethnographic sources in different libraries and research
centres. These have helped to clarify our view of the possible origin of the Tom\'ar\^aho and uncovered mentions of them in Jesuits' letters ({\it cartas annuas}) 
and old chronicles. Sequera, moreover,
lived among the natives for several periods between 1987 and 1992, which enabled him to carry out a methodical programme of recognition and transcription of the Tom\'ar\^aho
language, to compile a detailed inventory of the social representation of their flora and fauna, and to gather and register an immense mythical corpus that reveals the rich
world-view of this small group. The ethnological work combined the most pertinent qualitative techniques from anthropology with scientific rigour in observations. It also stressed the importance of fieldwork based on the method of participant observation, in line with the works of the Polish anthropologist Bronis{\l}aw Malinowski (Malinowski 1922).

The Tom\'ar\^aho (or Tomar\'axo) form a small ethnic group, which together with the Ybyt\'oso (or Ebid\'oso), makes up a larger group, the Ishir, known in Paraguay as the
Chamacoco. The Chamacoco, traditionally hunter-gatherers, are related linguistically to the Zamuco family. Another indigenous group, the Ayoreo, also
belongs to the Zamuco linguistic family. We do not propose to discuss the etymology of these words and names: disparate proposals abound, based mainly upon the writings
of European chroniclers, going back as far as {\it Viaje al R{\'\i}o de la Plata, 1534--1554} by harquebusier Ulrico Schmidl (1510--1580). In different catalogues of
languages and dialects, found within compilations of chronicles and other European texts, terms such as {\it Timinabas}, later {\it Timinaha}, and variants crop up,
always in the Zamuco category. It is mentioned that the peoples who spoke these languages lived in the Chaco woods, inland, far from the Paraguay river. It is also said that
these natives had not yet been `subdued by the Jesuit Mission'. We gather then that they are referring to the Tom\'ar\^aho, those natives whose descendants Sequera visited
in the neighbouring area of San Carlos in 1986 and who, due to their ancient customs and their own idiosyncrasies, remained alienated from Paraguayan society for many years.

Shamanism is just as relevant among the Chamacoco as it is in other indigenous cultures. Vocal music is closely related to rituals carried out by both male and female shamans. These prominent
members of the tribe, known as {\it konsaha} or {\it ahanak}, create their own repertoires based on dreams, called {\it chyk\^era}, which stimulate the creation
of their poems, melodies and rhythms. The Chamacoco shamans try to dominate their dreams by turning them into a chant called {\it teichu}. The production of these songs is
highly personal. While they can be transmitted to other members of the community, they are generally performed in groups where several songs mix together and a very distinctive musical
atmosphere is created, together with rattles ({\it sonajas}) and whistles as instrumental accompaniment. Rattles called {\it osecha} or {\it pa{\^\i}k\^ara} by the Chamacoco are made
from gourds ({\it calabazas} of the species {\it Lagenaria siceraria}) or with tortoise shells called {\it enermitak} ({\it Geochelone carbonaria}). Dry seeds or pebbles are
put inside so as to produce sound. These rattles represent the sky and the shamans identify their upper part with the centre of the sky, {\it porr hos\'ypyte}. Painted on the body of
the instrument are `visual narratives' in the shape of rhomboids, and inside these geometrical figures are representations of the stars, {\it porrebija} (Sequera
2006). Studies suggest a close relationship between the various Chamacoco vocal and instrumental techniques---especially among the Tom\'ar\^aho, where shamanism has lasted longer and
more intensively---and their vision of the world, a bond between musical expression and the natives' view of the universe that still needs to be researched in detail (see also Cordeu
1994). 

Chamacoco shamanic practices are very similar to those of other South American indigenous cultures, and even to those of other continents: visionary dreams, personality
split, trance achieved through chants, etc. are situations to be repeated in time and space. It is through these dreams that the {\it konsaha} discover for the
rest of the members of the tribe the true `topography' of the indigenous universe and the way it interrelates with the mythical tales inherited from their ancestors.\footnote{A
   detailed anthropological analysis of the case of the {\it Qom}, or {\it Tobas}, from the Argentinian Chaco can be consulted in Wright (2005).}

\vspace*{-0.5 cm}

\section{Astronomical elements of the Ybyt\'oso and Tom\'ar\^aho sky}

The Chamacoco imagine the world to be a disk-shaped flat surface which they call {\it h\~nymich}. Located on this immense earth disk are their familiar landscapes, their
villages, rivers and woods, and also the neighbouring villages they have contact with. The {\it h\~nymich} stands on the waters of an aquatic world called {\it niogorot urr}. As with other ancient peoples, the presence of a subterranean water world fits the Chamacoco world-view because of the importance they assign to water in springs and
rivers (moreover, the mythical beings called {\it ahnaps\^uro} are aquatic beings, as we shall see). The {\it niogorot urr} is subdivided into various strata at different depths. Above
the earth disk are situated several transparent skies, generically called {\it porrioho}. They are immense half spheres that surround men and which they imagine resting on
the sides of the {\it h\~nymich} disk.

The Chamacoco conception of the sky also includes many representations related to the
stars, individually as well as in groups. We find the Milky Way, called {\it iomyny} and meaning, perhaps, the way of the souls (Gim\'enez Ben{\'\i}tez {\it et al.} 2002),
and other `nebulae' clearly seen from the great Paraguayan Chaco, such as the Great ({\it kajywysta}) and Small ({\it kajywyhyrt\^a}) Magellanic Clouds. The Sun, called
{\it Deich}, and the Moon, {\it Xekulku}, both male characters, feature as protagonists in several stories that are very precious to the culture (Cordeu 1990--1991).  There are also certain
stories that mention Venus---which they call {\it Iohdle} and also mother of the stars, {\it porrebe bahlohta}---which are related to the other celestial bodies and to the
gentiles. Protectors are assigned to many elements of the Chamacoco cosmos: thus insects are protected by {\it \~N{\^\i}ogogo}, the frog {\it Bufo granulosus}, and certain local birds are protected by {\it Woh\^ora} (for the Tom\'ar\^aho) and by {\it P\^eeta yr\^ahata} (for the Ybyt\'oso)---`spirits' of the wilderness or of the forest. In the same way, stars are protected by {\it Abich}, the
star son of Venus, while the {\it \~nand\'u} or local ostrich ({\it Rhea Americana}), called {\it Pemme-Kamyt\^erehe} by the Chamacoco, is in charge of {\it Deich}, the Sun. The {\it kululte} or {\it ch\^aro} has also been represented in
Chamacoco stories and drawings. It is the `cosmic tree' or world support, which, as in many other cultures, functions
as a link between the {\it porrioho} and the {\it niogorot urr}. As we will see later, the {\it ch\^aro} plays a privileged role in stories concerning shamanic travels.

\vspace*{-0.5 cm}

\section{The axis mundi}

The centre exists as a symbol in many cultures and has been thoroughly studied by historians of religion. According to Eliade, the existence of a cosmic centre is a natural
consequence of the division of reality into the sacred (where all the value is concentrated) and the profane, whose space gives no orientation to man (Eliade 1998 [1957]). Thus the
world acquires meaning only through {\it hierophanies}.\footnote{from Greek, {\it hieros}, meaning sacred, and {\it phainein}, revelation, so that hierophany could be
  translated as `[where] the sacred is revealed'.} These intrusions of the sacred into the profane establish a unique place, a centre, which connects beyond the homogeneous
space unrelated to any mythical inheritance. It can also be interpreted as something that links several existential levels; among them, of course, is ordinary life,
but the other levels are inaccessible to men, or at least to most men.

The spatial arrangement of these different levels is oriented orthogonally to the flat space of the earth, although the borders of the habitable earth  also have
special connotations. Moving between these levels takes place vertically, either towards the upper part or, contrariwise, penetrating the bowels of the earth. The
typical picture that arises in the imagery of people from different cultures is that of a prominent mountain, which stands out in the landscape; or a `cosmic
tree' distinguished by its height or old age; or else some other type of `pillar' that functions as a link between the sky and the earth, as well as with the lower regions.

Generally speaking, this object is called an {\it axis mundi}, and history is full of examples.  We will see that there exists a strong parallel between these beliefs
and the cosmos as imagined by the Ybyt\'oso and the Tom\'ar\^aho from the northern Paraguayan Chaco, where the pillar will be represented by the {\it ch\^aro}.

The mountain or cosmic pillar was not only located in the centre of the organized space of ancient communities but its peak very often represented the highest point in the
world, an area that had not been reached even by the greatest floods ever. Such places were imagined, in their turn, to be a kind of navel of the earth, an embryo. The
Creation of the world took place there and then expanded towards the periphery in all directions. And of course, man had been born (had had his origin) in that centre of the
world; it was a centre of Creation.

\vspace*{-0.5 cm}

\section{The Ybyt\'oso and Tom\'ar\^aho axis mundi}

As we have said, in the Chamacoco view of the universe the world is supported by a tree called {\it kululte} or {\it ch\^aro}. This tree, which belongs to the
species {\it Chorisia insignis}, is endemic to Paraguay and bordering countries and is known in Spanish as {\it palo borracho} among other names. As in many other
cultures, this cosmic tree represents the link between the sky and the earth. It is said that in the roots of this mythological tree link together all graves.

The Chamacoco's upper universe is imagined to be a juxtaposition of transparent skies called {\it porrioho}, opposite to the dwelling of the
dead on earth where the {\it ch\^aro} sinks its roots. According to the Chamacoco origin myth, the earth and the sky were linked together by the cosmic tree in ancient times. While the two kingdoms were, so to speak, fused together, the inhabitants of the earth could move around without any barriers or impediments. The earliest of these inhabitants, called {\it yxyro poruwuhle}, fed
themselves without effort, hunting animals and gathering fruits easily. This situation is reminiscent of similar mythological eras in other cultures and it may very well be
seen as a sort of Chamacoco paradise or garden of abundance. However, as all the informants who collaborated with this ethnographic investigation stated, history
changed its course on the day when a widow called {\it Dagylta} and her children asked to be helped with food. No-one would give her any assistance. On seeing her neighbours' idleness and lack of concern, {\it Dagylta} turned into a beetle and slowly began to gnaw the wood of the mighty {\it ch\^aro}. At that moment a bird appears in the
story: {\it dichik{\^\i}or} of the species {\it Polyborus plancus} or {\it Caracara plancus}, known in Spanish as {\it carancho}. He attempted to stop the widow from
carrying out her plan, but he failed and the cosmic tree finally fell down. ({Among the {\it Mocov{\'\i}} of the Argentinian Chaco there exists a very similar story,
  where the widow is turned into a capybara or {\it carpincho}. This is documented in a 1764 chronicle by the Jesuit father Guevara (Guevara 1969).)

The two images shown in Fig. \ref{fig:balbuena} represent the conception of the cosmic tree as portrayed graphically by Ogwa Flores Balbuena, a member of the Ybyt\'oso community (Sequera
2005). The upper drawing, made in 1991, represents the Ybyt\'oso origin myth and shows the central place occupied by the {\it ch\^aro}. A variety of local birds and animals, together with a few people, are travelling between the earth and the sky. The lower drawing, made in 1988,
shows the tree {\it Ebyta} (another name for {\it ch\^aro}) as the pillar and support of the world, also during the time before its fall due to {\it Dagylta}'s
intervention. This drawing also shows the two kingdoms united by the mighty tree and some characters around it.

\begin{figure}[h] 
\begin{center}
\includegraphics[width=10.5cm]{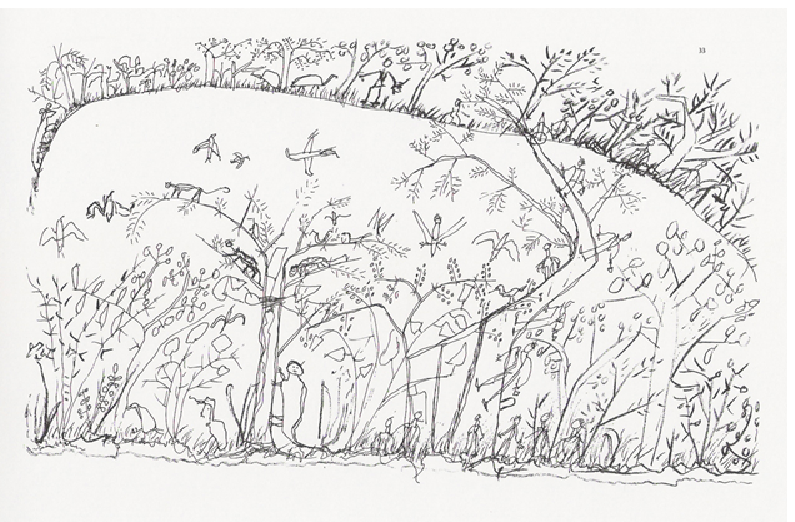}
\vspace*{0.5 cm}\\
\includegraphics[width=10.5cm]{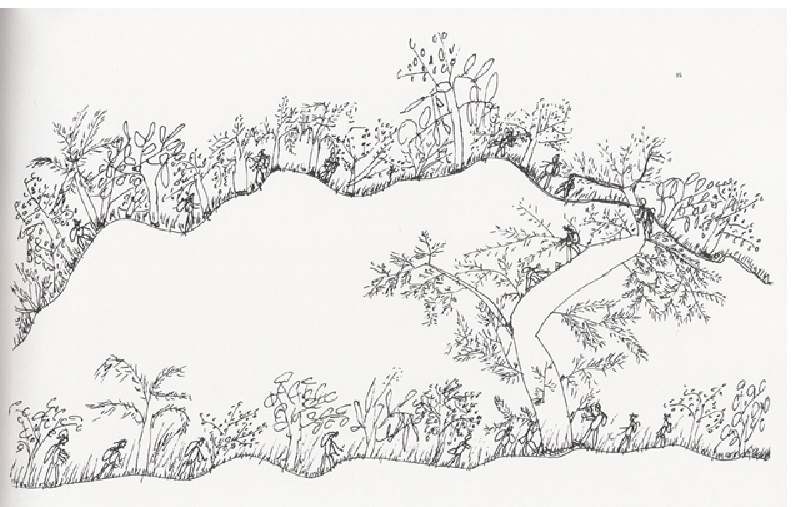}
\caption{Drawings by Ogwa Flores Balbuena, a member of the Ybyt\'oso \newline community, 1991 (top) and 1988 (bottom).
} 
\label{fig:balbuena}
\end{center}\end{figure}

When {\it Dagylta}, in the shape of a beetle, began gnawing and weakening the trunk of the cosmic tree, many of the people who until then had moved freely between the sky and the earth foresaw
what might happen and decided to climb down. Others, more idle, were left behind and once the tree had fallen, remained forever in the upper kingdom, clinging to the
sky. They turned into the {\it porrebija}, the stars that inhabit the Chamacoco sky. (Note that the Bushmen ({\it Bosquimanos}) of the Kalahari desert, South
  Africa, say that the stars are the very first peoples and that they are like them: nomads and hunter-gatherers (Krupp 1996). According to the Qom of the Argentinian Chaco,
  some people became stars following a cataclysm that altered the structure of earth and sky (Wright 2005).)

The Chamacoco say that when the bridge that joined the two kingdoms was cut off, the universe expanded and the sky and the earth never came together again. In
Ogwa Flores' representations, the sky is inhabited by beings that move about, interact and cohabit with animals and plants of the upper region---a region that is
located just a few metres above the tops of the highest trees in the terrestrial woods. Moreover, we gather from the stories that the colour of this same sky is reflected in the highest leaves of the woods. When {\it ch\^aro} collapsed, the two kingdoms became separated, transforming the sky, which was seen as a thick layer, hard and
grey, into a stratified region divided into several levels. The Chamacoco imagery, through this foundational event of the collapse, disconnects the universe into two opposing
worlds: the upper kingdom, which resembles a hemispherical sky, and the lower region, set on the original waters, which is imagined to be like a wrecked world (Sequera 2006).

This heaven-earth-netherworld separation has been extensively recorded both by anthropologists and historians of religions. Eliade (1974 [1955]) mentions the case of the Semang pygmies in the Malaca or Malayan peninsula. In the centre of the Semang world there is an enormous rock (or perhaps a limestone
hill; see Evans 1968 [1937]), called Batu Ribn or Batu 'Rem, which covers the lower regions and, in ancient times, was the base on which stood a tree trunk that reached up to the
sky. The netherworld, the centre of the earth and the entrance to heaven were joined by the same axis. This axis formed, in turn, the way to take if you wished to move from one region to another. As the story goes, communication with God and heaven was simple and natural for Semang people in the past, but after a ritual blunder this relationship was
ruptured. What had been natural for all the members of the Semang world was then left as a privilege only for the shamans.

\section{The organization of space in the Chamacoco world}

After the collapse of mythological {\it ch\^aro}, a catastrophic event that took place early in the history of the Ybyt\'oso and the Tom\'ar\^aho, the universe acquired a
particular architecture where the upper world is separated from the underground world. The former is located above ground and it includes
six strata, starting with the normal habitat of plants and animals. This layer, solid and dry, is also the dwelling place of men and rises to an altitude equal to the height of the highest
palm tree. This region is called {\it porr iut}, which in the Chamacoco language means lower sky. Next is a layer located above the ground, which is characterized by its
humidity. It is called {\it porr pehet}, which in the indigenous language signifies `halfway up to heaven' ({\it pehet} means space). It is within this layer that
clouds are located and rain develops. Storms, brought about by beings called {\it os\^asero} (spirits of the storms), also develop here. It is the
dwelling place of the wren-like rushbirds ({\it Phleocryptes melanops}; {\it pajaritos junqueros} in Spanish or {\it pet{\^\i}is chyperme} in the indigenous language), who fly low
in swamps, and it is also where some of the shamans dwell. Members of the Tom\'ar\^aho community affirm that this region is peopled by strange spirits called
{\it os{\^\i}oro kynaha}, evil beings that transmit deceases: microbial or contagious diseases are widespread in many regions of Northern Chaco. {\it Nehmurt}, a mythical malevolent being,
is also found in {\it porr pehet}.

The next stratum of the upper world is called {\it porr pixt} (true sky) and the Chamacoco represent it as a layer of thick fog, the domain of a character 
called {\it Lapyxe}, the rain maker. Many characters act as guardians of certain things, both phenomena and animate and inanimate objects, and {\it Lapyxe} is the guardian
of the waters. It is in this region that both the Moon and the Sun can be found but, according to some qualified informants, with the latter located beneath the former. In fact, the outline
of the Moon marks the upper edge of the {\it porr pixt}. This edge constitutes the gate to heaven; it is difficult to pass through as its guardians are  strange 
spirits ({\it os{\^\i}oro kynaha}). These are the main obstacles the shamans ({\it ahanak}) have to overcome in order to travel through the skies of the upper world during their
shamanic flights.

The fourth sky, {\it porr yhyr} (high sky), is a wide area where the stars ({\it porrebija}) are found, as well as groups of stars that make up asterisms and
constellations. Let us recall that in the Chamacoco world-view the stars came into existence as a result of the beings left behind in the upper realm at the moment when the {\it
  ch\^aro}, the cosmic tree that had joined the sky and the earth since time immemorial, collapsed. The most luminous `star' in the night sky after the Moon---the planet Venus, or {\it
  Iohdle} for the natives---is also found in this clear sky, together with other kinds of celestial objects. For example, the Milky Way, {\it iomyny}, which appears in the
crystal-clear skies of the Chaco Paraguayo as an outstanding whitish stripe across the sky, is also located in the {\it porr yhyr}. With the exception of a few particularly visionary
shamans, nobody has the power to enter this distant and profound sky.

The two last strata of the upper world are called {\it porr uhur} (`horizon' sky) and {\it porr nahnyk} (cold sky), respectively. The former is the
threshold to the end of the firmament and the region of the unknown. The latter is seen as an indefinite space which goes beyond the inner skies, the region where the
universe extends and the unknown predominates. It is a profound, airless sky.\footnote{Cordeu (1994) has provided a different stratification of the upper world, more
  related to the chromatic and atmospheric properties of the sky.} 

Let us travel now in the opposite vertical direction, i.e. towards the kingdom of the profound. The underground world, which, according to Chamacoco imagery, appeared after the
fall of the cosmic tree, is divided into three main strata. It is a hidden world that extends towards the bowels of the earth and has a viscose
constitution. It is supposed to be a region where destruction reigns. The first region, {\it n{\^\i}ogoro urr}, is a wetland zone with superficial water courses. This
fluid medium is inhabited by the mythical eel {\it dyhylygyta} and also the {\it uriche}, the present otter (of the species {\it Lontra longicaudis}). These animals coexist with
spirits that take the shape of fishes. It is these fishes who give the shamans powers to fight against strange spirits ({\it os{\^\i}oro kynaha}).

The second subterranean layer is imagined to be an area of deep waters mixed with thick viscous mud. It is known as {\it h\~nymich yhyrt}, literally `earth of the high hill',
and it is the dwelling place of monsters similar to eels. The most fabulous among them, both in terms of its size and its giant red head, is what the natives call {\it
  p\^eeta}. This monster eel grants to certain shamans the power to move quickly through this subterranean layer and to emerge from the water underworld at great speed at any
point on the earth. Finally, the third subterranean layer is called {\it h\~nymich urruo}, which means `under the ground'. This is the region of the dark and rotten world of
corpses and that of the being called {\it amyrmy lata}, which resembles a giant armadillo ({\it Priodontes maximus}) and which is connected with initiated shamans.

Although these different strata of the wrecked world cause destruction, death, and finally putrefaction of everything alive, they can nonetheless liberate certain forces of
ascension such as those that propel the shamans through the second layer (Sequera 2006) and also those that characterize the {\it ahnaps\^uro} (or {\it axn\'absero}), the mythical
beings of the Chamacoco world. Having been described for the first time by Guido Boggiani in 1900 
(Baldus 1932; Su\v{s}nik 1995 [1969]), 
the {\it ahnaps\^uro} are recollected by all Chamacoco natives, although only the
Tom\'ar\^aho maintain their ritual practices of mythical representation today (in the origin myths called {\it emuhno}). These powerful aquatic beings, whose bodies are
covered with scales and feathers, are believed to be the founders of Chamacoco culture. In times long ago, they lived together with the earliest inhabitants, {\it yxyro}, in full
harmony; they even taught the first peoples to search for food and organize themselves. But if they appear nowadays in the Tom\'ar\^aho camp this will cause shock and
terror. For this reason, rituals are organized regularly that will reduce the danger of their possible presence.

\section{Final discussion and future projects}

In this article we have given a brief sketch of the immense richness of Chamacoco conceptions of their relationship with nature and the sky. Painstaking ethnographic work has
brought to light the most remarkable features of this aboriginal group as a whole, and particularly the characteristics of the Tom\'ar\^aho and Ybyt\'oso. Through this work increasing numbers of people have become
familiar with their culture and come to develop a great respect for their traditions and cultural identity. The transcription of their language, and their slow social improvement and education, are some of the elements that will help others to build upon this work in the future.

In the course of this vital ethnographic work, however, astronomical knowledge and practices have not been treated as thoroughly as the other aspects. Among the things that remain to be
done are a detailed analysis of the recognition and identification of outstanding aspects of the Chamacoco sky, and an in-depth study of the true meaning of the {\it
  ch\^aro}, the cosmic tree that once united the sky and the earth and which remains, even now, central to rituals concerning the origin of the world. Other aspects needing greater attention are the names of, and the stories that were almost certainly told about, prominent stars visible
at different times of the year, such as Sirius, and prominent asterisms such as Orion's belt; imagery concerning the presence and characteristics of the Milky Way; and their
interpretation of certain exceptional and sudden phenomena, such as total solar eclipses, and of objects that make prolonged appearances in the sky, such as comets, or sporadic ones such as meteors.

The story of {\it Iodhle} (Venus), who long ago married a young Tom\'ar\^aho, is just one of many stories with astronomical elements that old people used to
tell. Many of these stories are still remembered by older members of this group. It is necessary and urgent to try to incorporate aspects of this culture into the intangible
heritage of humanity before it is lost forever. Another project being planned is to try to understand the relationship between the vocal and
instrumental techniques of the Chamacoco and their view of the universe surrounding them, especially in the use they make of rattles ({\it pa{\^\i}k\^ara}) 
which, as we have seen, represent the starry sky.

As part of our project we intend to work together with members of the present Tom\'ar\^aho community to gather data on the recognition of stars and
constellations and their distribution, dark zones in the sky, nebulae etc. These data would be presented as annotated maps that would convey the symbolic significance of the Tom\'ar\^aho sky of the Upper Chaco
region at different times of the year. This could tell us a great deal about aboriginal imagery and also about many aspects of their everyday life that they project into
the dark depths of the sky. In short, we consider that the Tom\'ar\^aho conception of the sky has not been sufficiently explored as a study in itself, and that more fieldwork
would be welcome in the area of ethnoastronomy as an interdisciplinary activity that includes both anthropologists and astronomers. 

\begin{acknowledgments}
We wish to thank Beatriz Tosso, whose key input benefitted the final shape of this article. A. G. acknowledges support from CONICET and from the University of Buenos Aires. Both authors would like to thank their colleagues participating in the {\it Paran\'a Ra'anga} project (AECID), where this collaboration started, for many fruitful discussions.
\end{acknowledgments}

\end{document}